\documentclass[aps,preprint]{revtex4}%
\usepackage{amsfonts}
\usepackage{amsmath}
\usepackage{amssymb}
\usepackage{graphicx}%
\begin{document}
\title{Spin diffusion in an inhomogeneous internal field (non equidistant energy spectrum)}
\author{Gregory B. Furman and Shaul D. Goren}
\pacs{PACS number}

\begin{abstract}
The theory of NQR spin diffusion is extended to the case of spin lattice
relaxation and spin diffusion in an inhomogeneous field. Two coupled equations
describing the mutual relaxation and the spin diffusion of the nuclear
magnetization and dipolar energy were obtained by using the method of
nonequilibrium state operator. The equations were solved for short and long
times approximation corresponding to the direct and diffusion relaxation regimes.

\end{abstract}
\maketitle

\textbf{Introduction}

Studies of the NMR \cite{bloum,khu57,pdg58} and NQR of nuclei have
demonstrated that spin diffusion plays an important role in the relaxation of
nuclei in the presence of paramagnetic impurities (PI). Such type of
relaxation originates from the magnetic dipole-dipole interaction of PI with
neighboring nuclei, which is inversely proportional to the sixth power of the
distance. Thus, near the PI the equilibrium with the lattice is reached at a
faster rate \cite{bloum,khu57,pdg58}. The nuclear magnetization during the
relaxation process is spatially inhomogeneous over a sample volume, and this
induces a spatial diffusion of the nuclear spin energy by, for example,
flip-flop transitions due to the dipole-dipole interactions between nuclear spin.

However, first, till now most studies of the nuclear spin diffusion were
related to systems with nuclear spin $I=1/2$, described by the Hamiltonian,
whose main term includes just linear functions of spin operators and,
correspondingly, forms equidistant energy spectra and, second, most of them
deal with the process of a spin diffusion in homogeneous magnetic fields
\cite{khu66,khu69}.

In many samples spin systems consist of nuclei with $I>1/2$ and they interact
with their environment through the electric quadrupole moment $Q$, and these
interactions are strong enough to observe magnetic resonance of nuclei in the
absence of an external magnetic field (pure NQR-case). Unlike the NMR-case,
the NQR energy spectrum is non-equidistant and, in many cases, degenerated.
These circumstances lead to certain difficulties in obtaining a diffusion
equation and a calculation of a diffusion coefficient.

Bisides, the spin diffusion processes no longer exactly conserve nuclear
quadrupole energy in an inhomogeneous field \cite{GenackRedfield1973} because
the quadrupole interaction energy is not identical for neighboring nuclear
spins. In order for the spin diffusion process to take place, the nuclear
quadrupole energy difference must be taken up by another thermodynamic
reservoir, for example, by the dipole-dipole energy one.

Recently the theories for spin diffusion of the nuclear dipolar order via PI's
\cite{fur1999,Greenbaum} in NMR and NQR \cite{Furman2000} and spin diffusion
in an inhomogenius feild \cite{Furman2004} have been developed. It was shown
that thermodynamics resevoir of the dipolar order plays an importent role in
the spin latice relaxation and spin diffusion in an inhomogenius feild.
Nuclear dipolar order is characterized by a state with nuclear spins oriented
along an internal local field generated by the dipole-dipole interactions
(DDI) and can be described by a dipolar temperature
\cite{jeneer,gold,abrg,slic}. Here we consider the phenomena of spin lattice
relaxation and spin diffusion for both the nuclear quadrupole and dipolar
energies of the nuclear spins due to their DDI in solids containing PI's in an
inhomogeneous field.

\textbf{Theory}

1. \textbf{Hamiltonian}

The evolution of the spin system consisting of nuclear spins with $I>1/2$ and
PI spins may be described by a solution of the equation for density matrix
$\rho\left(  t\right)  $ $\left(  \text{in units of }\hbar=1\right)  $%

\begin{equation}
i\frac{d\rho\left(  t\right)  }{dt}=\left[  \mathcal{H}\left(  t\right)
,\rho\left(  t\right)  \right]  \tag{1}%
\end{equation}
with the Hamiltonian%

\begin{equation}
\mathcal{H}(t)=\mathcal{H}_{Q}+\mathcal{H}_{dd}+\mathcal{H}_{PI}%
+\mathcal{H}_{P}+\mathcal{H}_{br}\left(  t\right)  . \tag{2}%
\end{equation}
Here $\mathcal{H}_{Q}$ represents the interaction of the $I$-spin system with
the EFG; $\mathcal{H}_{dd}$ and $\mathcal{H}_{PI}$ are the Hamiltonians of the
dipole-dipole interaction between nuclear spins and nuclear and PI spins,
respectively; $\mathcal{H}_{P}$ describes the impurity spin system;
$\mathcal{H}_{br}\left(  t\right)  =\sum_{q=-2}^{2}E^{\left(  -q\right)
}\left(  t\right)  A^{q}$ the spin-lattice interaction Hamiltonian , describes
spin-lattice relaxation caused by the torsional vibrations (Bayer mechanism)
\cite{bayer6}, where $A^{q}$ is a bilinear function of the spin operators and
$E^{\left(  -q\right)  }\left(  t\right)  $ is a random function of time
\cite{abrg}.

Using the projection operators \cite{Furman1983} $e_{mn}^{\mu}$ and
$\varepsilon_{mn}^{j}$ defined by their matrix elements $\left\langle
m^{^{\prime}}\left\vert e_{mn}^{\mu}\right\vert n^{^{\prime}}\right\rangle
=\delta_{m^{^{\prime}}m}\delta_{n^{^{\prime}}n}$ and $\left\langle
\nu^{^{\prime}}\left\vert \varepsilon_{\nu\sigma}^{\mu}\right\vert
\sigma^{^{\prime}}\right\rangle =\delta_{\nu^{^{\prime}}\nu}\delta
_{\sigma^{^{\prime}}\sigma}$ and introducing a projection density operators,
$e_{mn}\left(  \vec{r}\right)  $, for the nuclear spins $I$, and
$\varepsilon_{mn}\left(  \vec{r}\right)  $ for PI spins%

\begin{equation}
e_{mn}\left(  \vec{r}\right)  =\sum_{\mu}\delta\left(  \vec{r}-\vec{r}_{\mu
}\right)  e_{mn}^{\mu};\text{ }\varepsilon_{mn}\left(  \vec{r}\right)
=\sum_{j}\delta\left(  \vec{r}-\vec{r}_{j}\right)  \varepsilon_{mn}^{j}
\tag{3}%
\end{equation}
the density of the Hamiltonians $\mathcal{H}_{Q}$, $\mathcal{H}_{dd}$ , and
$\mathcal{H}_{PI}$ can be written down in the following form:%

\begin{equation}
\mathcal{H}_{Q}\left(  \vec{r}\right)  =\left(  2I+1\right)  ^{-1}\sum
_{mn}\omega_{mn}^{0}e_{mm}\left(  \vec{r}\right)  , \tag{4}%
\end{equation}

\begin{equation}
\mathcal{H}_{dd}\left(  \vec{r}\right)  =\int d\vec{r}^{\prime}\sum
_{mnm^{^{\prime}}n^{^{\prime}}}g_{mn}^{m^{^{\prime}}n^{^{\prime}}}\left(
\vec{r}-\vec{r}^{\prime}\right)  e_{mn}\left(  \vec{r}\right)  e_{m^{^{\prime
}}n^{^{\prime}}}\left(  \vec{r}^{\prime}\right)  , \tag{5}%
\end{equation}

\begin{equation}
\mathcal{H}_{PI}\left(  \vec{r}\right)  =\int d\vec{r}^{\prime}\sum
_{mnm^{^{\prime}}n^{^{\prime}}}f_{mn}^{m^{^{\prime}}n^{^{\prime}}}\left(
\vec{r}-\vec{r}^{\prime}\right)  e_{mn}\left(  \vec{r}\right)  \varepsilon
_{m^{^{\prime}}n^{^{\prime}}}\left(  \vec{r}^{\prime}\right)  , \tag{6}%
\end{equation}%
\begin{equation}
\mathcal{H}_{br}\left(  t\right)  =\sum_{q}\sum_{mn}E^{\left(  -q\right)
}\left(  t\right)  A_{mn}^{q}\int d\vec{r}e_{mn}\left(  \vec{r}\right)
\tag{7}%
\end{equation}
where $\omega_{mn}^{0}=\lambda_{m}-\lambda_{n}$, $\lambda_{m}$, $\left\vert
m\right\rangle ,$ and $\left\vert n\right\rangle $ are the eigenvalues and
eigenvectors of the operator $\mathcal{H}_{Q}$. $\left\vert \nu\right\rangle $
and $\left\vert \sigma\right\rangle $ are eigenvectors of the operator
$\mathcal{H}_{P}$; matrix elements $g_{mn}^{m^{^{\prime}}n^{^{\prime}}}\left(
\vec{r}-\vec{r}^{\prime}\right)  $ and $f_{mn}^{m^{^{\prime}}n^{^{\prime}}%
}\left(  \vec{r}-\vec{r}^{\prime}\right)  $ can be presented as%

\begin{equation}
g_{mn}^{m^{^{\prime}}n^{^{\prime}}}\left(  \vec{r}-\vec{r}^{\prime}\right)
=G_{mn}^{m^{^{\prime}}n^{^{\prime}}}\left(  \vec{r}-\vec{r}^{\prime}\right)
\left[  \left(  \delta_{mn}+\delta_{pq}\right)  \left(  \delta_{m\bar{n}%
}+\delta_{p\bar{q}}\right)  +\left(  \delta_{mq}+\delta_{pn}\right)  \left(
\delta_{m\bar{q}}+\delta_{p\bar{n}}\right)  \right]  \tag{8}%
\end{equation}
with $\bar{n}=n$ and%

\begin{equation}
f_{mn}^{m^{^{\prime}}n^{^{\prime}}}\left(  \vec{r}-\vec{r}^{\prime}\right)
=F_{mn}^{m^{^{\prime}}n^{^{\prime}}}\left(  \vec{r}-\vec{r}^{\prime}\right)
\left(  \delta_{mn}+\delta_{pq}\right)  \left(  \delta_{m\bar{n}}%
+\delta_{p\bar{q}}\right)  , \tag{9}%
\end{equation}
$G_{mn}^{m^{^{\prime}}n^{^{\prime}}}$ and $F_{mn}^{m^{^{\prime}}n^{^{\prime}}%
}$ are matrix elements of the dipole-dipole Hamiltonians $\mathcal{H}_{dd}$
and $\mathcal{H}_{PI}$ in $\mathcal{H}_{Q}$-representation \cite{Furman1983}.

2. \textbf{Diffusion equations}

To obtain the equation describing the spin diffusion and spin lattice
relaxation of both the quadrupole and dipolar orders we will use the method of
nonequilibrium state operator \cite{zub}, which has been applied to obtain the
diffusion equation in cases of the Zeeman order spin diffusion \cite{buizub}
and dipolar order \cite{fur1999} spin diffusion.

Using the commutation rules between the components of the projection operators
(3) $\left[  e_{mn}\left(  \vec{r}\right)  ,e_{m^{\prime}n^{\prime}}\left(
\vec{r}^{\prime}\right)  \right]  =\delta\left(  \vec{r}-\vec{r}^{\prime
}\right)  \left(  \delta_{nm^{\prime}}e_{mn^{\prime}}\left(  \vec{r}\right)
-\delta_{n^{\prime}m}e_{m^{\prime}n}\left(  \vec{r}\right)  \right)  $ and
$\left[  e_{mn}\left(  \vec{r}\right)  ,\varepsilon_{m^{^{\prime}}n^{^{\prime
}}}\left(  \vec{r}^{\prime}\right)  \right]  =0$, we can obtain the following
equations in the form of localized laws of conservation of the spin energy densities%

\begin{equation}
\frac{\partial e_{mm}\left(  \vec{r}\right)  }{\partial t}+div\left(  \vec
{j}_{mm}\left(  \vec{r}\right)  \right)  =K_{mm}\left(  \vec{r}\right)
+L_{mm}\left(  \vec{r}\right)  , \tag{10}%
\end{equation}

\begin{equation}
\frac{\partial\mathcal{H}_{dd}\left(  \vec{r}\right)  }{\partial t}+div\left(
\vec{j}_{dd}\left(  \vec{r}\right)  \right)  +\sum_{mn}\vec{j}_{mm}\left(
\vec{r}\right)  \frac{\partial\omega_{mn}^{0}\left(  \vec{r}\right)
}{\partial\vec{r}}=K_{dP}\left(  \vec{r}\right)  +L_{dP}\left(  \vec
{r}\right)  , \tag{11}%
\end{equation}%
\begin{equation}
\frac{\partial\mathcal{H}_{P}}{\partial t}=-\int d\vec{r}\left\{  \left(
2I+1\right)  ^{-1}\frac{\partial}{\partial t}\left[  \sum_{mn}\beta
_{mn}\left(  \vec{r}\right)  \omega_{mn}^{0}\left(  \vec{r}\right)
e_{mm}\left(  \vec{r}\right)  \right]  +\frac{\partial\mathcal{H}_{dd}\left(
\vec{r}\right)  }{\partial t}\right\}  . \tag{12}%
\end{equation}
The last equation is result of the energy conservation law. In Eqs (10)
$\vec{j}_{mm}\left(  \vec{r}\right)  $ is the flux of operator $e_{mm}\left(
\vec{r}\right)  $,%

\begin{align}
\vec{j}_{mm}\left(  \vec{r}\right)   &  =-\frac{i}{2}\int d\vec{r}^{\prime
}\sum_{k}\left(  \vec{r}-\vec{r}^{\prime}\right)  G_{mk}^{km}\left(  \vec
{r}-\vec{r}^{\prime}\right)  \{e_{mk}\left(  \vec{r}^{\prime}\right)
e_{km}\left(  \vec{r}\right)  -e_{mk}\left(  \vec{r}\right)  e_{km}\left(
\vec{r}^{\prime}\right) \nonumber\\
&  -e_{km}\left(  \vec{r}^{\prime}\right)  e_{mk}\left(  \vec{r}\right)
+e_{km}\left(  \vec{r}\right)  e_{mk}\left(  \vec{r}^{\prime}\right)  \}
\tag{13}%
\end{align}
$K_{mm}\left(  \vec{r}\right)  =i[\mathcal{H}_{PI},e_{mm}\left(  \vec
{r}\right)  ]$ and \ $L_{mm}\left(  \vec{r}\right)  =i[\mathcal{H}_{br}%
,e_{mm}\left(  \vec{r}\right)  ]$ in Eq. (10) are the change of the nuclear
quadrupolar energy density due to the interaction with the PI and caused by
the torsional vibrations (Bayer mechanism) , respectively. In Eq.(11),
$\vec{j}_{dd}\left(  \vec{r}\right)  $ is the operator of the flux of nuclear
dipolar energy,%

\begin{align}
\vec{j}_{dd}\left(  \vec{r}\right)   &  =-2i\sum_{mnpq}\sum_{m^{^{\prime}%
}n^{^{\prime}}p^{^{\prime}}}\int d\vec{r}^{^{\prime}}\int d\vec{r}%
^{^{^{\prime\prime}}}\left(  r^{^{\prime\prime}}-r\right)  G_{mnpq}\left(
r^{^{\prime\prime}}-r\right)  e_{mn}\left(  \vec{r}^{^{\prime\prime}}\right)
e_{m^{^{\prime}}n^{^{\prime}}}\left(  \vec{r}^{^{\prime}}\right) \nonumber\\
&  \left[  G_{m^{^{\prime}}n^{^{\prime}}qp^{^{\prime}}}\left(  r^{^{\prime}%
}-r\right)  e_{pp^{^{\prime}}}\left(  \vec{r}\right)  -G_{m^{^{\prime}%
}n^{^{\prime}}p^{^{\prime}}p}\left(  r^{^{\prime}}-r\right)  e_{p^{^{\prime}%
}q}\left(  \vec{r}\right)  \right]  , \tag{14}%
\end{align}
$K_{dP}\left(  \vec{r}\right)  =i\left[  \mathcal{H}_{PI},\mathcal{H}%
_{dd}\left(  \vec{r}\right)  \right]  $ and \ $L_{dP}\left(  \vec{r}\right)
=i\left[  \mathcal{H}_{br},\mathcal{H}_{dd}\left(  \vec{r}\right)  \right]  $
in Eq. (11) are the change of the nuclear dipolar energy density due to the
interaction with the PI and caused by the torsional vibrations (Bayer
mechanism) , respectively. Note, that in the case with a homogenous magnetic
field, $\frac{\partial\omega_{mn}^{0}\left(  \vec{r}\right)  }{\partial\vec
{r}}$, from the system equations (10) \ and (11) we have two separate
equations: Eq. (10) leads to the localized law of conservation of the
quadrupolar energy densities \cite{Furman2000a} and Eq. (11) leads to
conservation law of the dipolar energy \cite{Furman2000}.

In the high temperature approximation we can write the density matrix in the
following form \cite{zub}%

\begin{equation}
\sigma=\left\{  1-\int_{0}^{1}d\lambda\left[  \mathcal{B}\left(
t+i\lambda\right)  -\left\langle \mathcal{B}\left(  t+i\lambda\right)
\right\rangle \right]  \right\}  \sigma_{eq}, \tag{15}%
\end{equation}
where the thermodynamic average $\left\langle ...\right\rangle $ corresponds
to an average with the quasi-equilibrium operator $\sigma_{eq}$%
=$e^{-\mathcal{A}}/Tre^{-\mathcal{A}}$, and%

\begin{equation}
\mathcal{A}=\int d\vec{r}\left[  \left(  2I+1\right)  ^{-1}\sum_{mn}\beta
_{mn}\left(  \vec{r}\right)  \omega_{mn}^{0}\left(  \vec{r}\right)
e_{mm}\left(  \vec{r}\right)  +\beta_{d}\left(  \vec{r}\right)  \mathcal{H}%
_{dd}\left(  \vec{r}\right)  \right]  -\beta_{p}\mathcal{H}_{P}, \tag{16}%
\end{equation}

\begin{gather}
\mathcal{B}\left(  t+i\lambda\right)  =e^{-\lambda\mathcal{A}}\int_{-\infty
}^{0}dte^{\varepsilon t}\int d\vec{r}\left(  2I+1\right)  ^{-1}\sum
_{mn}\{\omega_{mn}^{0}\left(  \vec{r}\right)  \vec{j}_{mm}\left(  \vec
{r},t\right)  \nabla\beta_{mn}\left(  \vec{r},t\right)  +\vec{j}_{mm}\left(
\vec{r},t\right)  \times\nonumber\\
\times\left[  \beta_{mn}\left(  \vec{r},t\right)  -\beta_{d}\left(  \vec
{r},t\right)  \right]  \frac{\partial\omega_{mn}^{0}\left(  \vec{r}\right)
}{\partial\vec{r}}+\vec{j}_{d}\left(  \vec{r},t\right)  \nabla\beta_{d}\left(
\vec{r},t\right) \nonumber\\
+\left[  \beta_{mn}\left(  \vec{r},t\right)  -\beta_{L}\right]  K_{ZS}\left(
\vec{r},t\right)  +\left[  \beta_{d}\left(  \vec{r},t\right)  -\beta
_{L}\right]  K_{dS}\left(  \vec{r},t\right)  \}e^{\lambda\mathcal{A}}.
\tag{17}%
\end{gather}
By using Eqs. (10) -(16), and taking into account that for single crystal
sample of cubic symmetry, the diffusion coefficients, both for quadrupolar and
dipolar energies, which in the general case of noncubic symmetry is a
symmetrical tensor of second rank \cite{khu69}, reduces to a scalar quantity.
By introduction the quantities $\left[  \beta_{mn}\left(  \vec{r},t\right)
-\beta_{L}\right]  =\xi_{mn}\left(  \vec{r},t\right)  $ $\ $and $\left[
\beta_{d}\left(  \vec{r},t\right)  -\beta_{L}\right]  =\zeta\left(  \vec
{r},t\right)  $, the diffusion equations can be obtained%
\begin{align}
\frac{\partial\xi_{mn}\left(  \vec{r},t\right)  }{\partial t}  &  =\frac
{1}{\omega_{mn}^{0}\left(  \vec{r}\right)  }\nabla\left\{  D_{mn}\left(
\vec{r}\right)  [\omega_{mn}^{0}\left(  \vec{r}\right)  \nabla\xi_{mn}\left(
\vec{r},t\right)  +\left(  \xi_{mn}\left(  \vec{r},t\right)  -\zeta\left(
\vec{r},t\right)  \right)  \nabla\omega_{mn}^{0}\left(  \vec{r}\right)
]\right\}  -\nonumber\\
&  -W_{mn}\left(  \vec{r}\right)  \xi\left(  \vec{r},t\right)  , \tag{18}%
\end{align}%
\begin{align}
\frac{\partial\zeta\left(  \vec{r},t\right)  }{\partial t}  &  =\sum_{mn}%
\frac{D_{mn}\left(  \vec{r}\right)  \nabla\omega_{mn}^{0}\left(  \vec
{r}\right)  }{M_{mm}}[\omega_{mn}^{0}\left(  \vec{r}\right)  \nabla\xi
_{mn}\left(  \vec{r},t\right)  +\left(  \xi_{mn}\left(  \vec{r},t\right)
-\zeta\left(  \vec{r},t\right)  \right)  \nabla\omega_{mn}^{0}\left(  \vec
{r}\right)  ]+\nonumber\\
&  +\nabla\left[  D_{d}\left(  \vec{r}\right)  \nabla\zeta\left(  \vec
{r},t\right)  \right]  -W_{d}\left(  \vec{r}\right)  \zeta\left(  \vec
{r},t\right)  . \tag{19}%
\end{align}
where $M_{mm}=\int d\vec{r}^{\prime}G^{2}\left(  \vec{r}-\vec{r}^{\prime
}\right)  \left\langle e_{mm}^{2}\left(  \vec{r}\right)  \right\rangle $.

The first term in the square brackets of the right side of Eq. (18) describes
the time dependence of the $\xi_{mn}\left(  \vec{r},t\right)  $ due to the
spin diffusion with a diffusion coefficient of%

\begin{equation}
D_{mn}\left(  \vec{r}\right)  =\int_{0}^{1}d\lambda\int_{-\infty}^{\infty
}dte^{\varepsilon t}\left\langle \vec{j}_{mn}\left(  \vec{r},\lambda,t\right)
\vec{j}_{mn}\left(  \vec{r}\right)  \right\rangle /Tr\left(  e_{mm}^{2}\left(
\vec{r}\right)  \right)  \tag{20}%
\end{equation}
The second term gives the variation of $\beta_{mn}\left(  \vec{r},t\right)  $
due to interaction with the dipolar reservoir in the inhomogeneous field. The
last term in the right side of Eq. (18) gives the relaxation of $\xi
_{mn}\left(  \vec{r},t\right)  $ toward the inverse lattice temperature with
density of the transition probability per unit time, $W_{mn}\left(  \vec
{r}\right)  ,$which for a cubic crystal is given by $\left(  K_{mm}\left(
\vec{r}\right)  +L_{mm}\left(  \vec{r}\right)  \right)  $, where
\begin{equation}
W_{mn}\left(  \vec{r}\right)  =\int_{0}^{1}d\lambda\int_{-\infty}^{\infty
}dte^{\varepsilon t}\left\langle \left(  K_{mm}\left(  \vec{r},\lambda
,t\right)  +L_{mm}\left(  \vec{r},\lambda,t\right)  \right)  \left(
K_{mm}\left(  \vec{r}\right)  +L_{mm}\left(  \vec{r}\right)  \right)
\right\rangle /Tr\left(  e_{mm}e_{mm}\left(  \vec{r}\right)  \right)  \tag{21}%
\end{equation}
The first term in the curly brackets of right side of Eq. (19) describes the
time variation of the dipolar energy due to the spin diffusion with diffusion coefficient%

\begin{equation}
D_{d}\left(  \vec{r}\right)  =\int_{0}^{1}d\lambda\int_{-\infty}^{\infty
}dte^{\varepsilon t}\left\langle \vec{j}_{d}\left(  \vec{r},\lambda,t\right)
\vec{j}_{d}\left(  \vec{r}\right)  \right\rangle /Tr\left(  \mathcal{H}%
_{d}\mathcal{H}_{d}\left(  \vec{r}\right)  \right)  . \tag{22}%
\end{equation}
The second term gives the variation of $\beta_{d}\left(  \vec{r},t\right)  $
due to the interaction with the quadrupole reservoir in an inhomogeneous
field. The last term in the right side of Eq. (19) gives the relaxation with
the density of the transition probability per unit time, $W_{d}\left(  \vec
{r}\right)  $, which for a cubic crystal is given by
\begin{equation}
W_{d}\left(  \vec{r}\right)  =\int_{0}^{1}d\lambda\int_{-\infty}^{\infty
}dte^{\varepsilon t}\left\langle \left[  K_{dP}\left(  \vec{r},\lambda
,t\right)  +L_{dP}\left(  \vec{r},\lambda,t\right)  \right]  \left[
K_{dP}\left(  \vec{r}\right)  +L_{dP}\left(  \vec{r}\right)  \right]
\right\rangle /Tr\left(  \mathcal{H}_{d}\mathcal{H}_{d}\left(  \vec{r}\right)
\right)  \tag{23}%
\end{equation}

The boundary conditions can be introduced by defining a sphere with radius $l$
about each PI, called the spin diffusion barrier radius.{\Large \ }Inside this
sphere the spin diffusion process goes to zero:%
\begin{equation}
\nabla\beta_{mn}\left(  \vec{r},t\right)  \mid_{\left\vert \vec{r}\right\vert
=l}=0\text{ and }\nabla\beta_{d}\left(  \vec{r},t\right)  \mid_{\left\vert
\vec{r}\right\vert =l_{d}}=0. \tag{24}%
\end{equation}
To obtain the radius of spin diffusion barrier, let us emphasize that
$D_{mn}\left(  \vec{r}\right)  $ and $D_{d}\left(  \vec{r}\right)  $\ are a
function of the distance $\vec{r}$ from the nearest PI. In the Gaussian
limiting case the stochastic theory of magnetic resonance \cite{Kubo} the
dimensional dependence of the diffusion coefficients $D_{mn}\left(  \vec
{r}\right)  $ and $D_{d}\left(  \vec{r}\right)  $ can be expressed by the
following function \cite{buizub}%

\begin{equation}
D_{mn}\left(  \vec{r}\right)  ,D_{d}\left(  \vec{r}\right)  \sim\exp\left[
-\left(  \frac{\nabla\omega_{mn}^{0}\left(  \vec{r}\right)  \vec{r}}%
{\omega_{d}}\right)  ^{2}\right]  . \tag{25}%
\end{equation}
Using Eq. (26) the diffusion barrier radius \cite{bloum,khu57,blum,Ror,kes}
for the spin diffusion of quadrupole energy can be found by solving the
equation,{\Large \ }%
\begin{equation}
\frac{3\gamma_{S}}{l^{3}}\left\langle S_{z}\right\rangle \left[  \frac{qe\mu
}{\left\langle S_{z}\right\rangle \gamma_{S}}\left\vert \vec{r}_{0}\right\vert
+1\pm\left(  1-\frac{\left\vert \vec{r}_{0}\right\vert }{l}\right)  \right]
=\frac{6\gamma_{I}}{\left\vert \vec{r}_{0}\right\vert ^{3}}, \tag{26}%
\end{equation}
{\Large \ }where the first term in square brackets in Eq. (26) describes a
distortions of the crystal field as a result of the inclusion of the PI. In
Eq.(26) $r_{0}$\ is distance between neighboring nuclei, $\mu$\ is the
Sternheimer antishielding factor \cite{das}. It was assumed that the
distortion of the electric field is equivalent to the presence of a charge
$q$\ \cite{kes}$.$

The examination of the functional dependence (25) for the dipolar diffusion
coefficient results that the main term for $D_{d}\left(  \vec{r}\right)  $
does not include any dimensional dependence. Thus the radius of the diffusion
barrier for dipolar energy is $l_{d}=\left\vert \vec{r}_{0}\right\vert $ ,
which corresponds to non-barrier diffusion and to the fastest relaxation of
the dipolar energy.

In the case of a homogeneous magnetic field, $\nabla\omega_{mn}^{0}\left(
\vec{r}\right)  =0$, equations (20) and (21) give the results obtain earlier
for the spin diffusion of the quadrupole \cite{khu69} and of the dipolar
energies \cite{fur1999}. From Eqs. (20) and (21) we get that the dissipation
of the density quadrupolar and dipolar energies are driven by: i) the exchange
between them; ii) spin diffusion process: and iii) direct relaxation to the PI.

\bigskip

2. \textbf{Direct relaxation regime}

\bigskip

Exact solutions of Eqs. (20) and (21) are extremely difficult, even for simple
a model situations. That is why we consider evolution of the spin system in
time by using the next considerations. Immediately after a disturbance of the
nuclear spin system, the gradients of $\xi_{mn}\left(  \vec{r},t\right)  $ and
$\zeta\left(  \vec{r},t\right)  $ are sufficiently small and diffusion cannot
be of importance at the start of the relaxation process \cite{blum}, this is
the so called diffusion vanishing regime \cite{Lowe}. To describe the
relaxation at that time interval we can use Eqs.(20) and (21) by putting all
inverse temperature gradient-terms equal zero, $\nabla\xi_{mn}\left(  \vec
{r},t\right)  =0$ and $\nabla\zeta\left(  \vec{r},t\right)  =0$. We also
accept the approximation that at distances larger then the radius of the
diffusion barrier the diffusion coefficient is independent of $\vec{r}$
\cite{khu66}. Under these approximations Eqs. (20) and (21) come to%
\begin{equation}
\frac{\partial\xi_{mn}\left(  \vec{r},t\right)  }{\partial t}=\frac
{D_{mn}\left(  \vec{r}\right)  \Delta\omega_{mn}^{0}\left(  \vec{r}\right)
}{\omega_{mn}^{0}\left(  \vec{r}\right)  }\left\{  [\left(  \xi_{mn}\left(
\vec{r},t\right)  -\zeta\left(  \vec{r},t\right)  \right)  ]\right\}
-W_{mn}\left(  \vec{r}\right)  \xi\left(  \vec{r},t\right)  , \tag{27}%
\end{equation}%
\begin{equation}
\frac{\partial\zeta\left(  \vec{r},t\right)  }{\partial t}=\sum_{mn}%
\frac{D_{mn}\left(  \vec{r}\right)  \left[  \nabla\omega_{mn}^{0}\left(
\vec{r}\right)  \right]  ^{2}}{M_{mm}}[\xi_{mn}\left(  \vec{r},t\right)
-\zeta\left(  \vec{r},t\right)  ]-W_{d}\left(  \vec{r}\right)  \zeta\left(
\vec{r},t\right)  . \tag{28}%
\end{equation}
The evolutions of the $\xi_{mn}\left(  \vec{r},t\right)  $ and $\zeta(\vec
{r},t)$ toward their steady-state values is a linear combination of $(2I+1)$
exponents with $N=(2I+1)$ relaxation times $\tau_{N}\left(  \vec{r}\right)  $.
These relaxation times $\tau_{N}\left(  \vec{r}\right)  $ are the function of
the position $\vec{r}$. In order to obtain the experimentally observed
signals, the solutions of Eq. (27) and (28),$\ $ must be averaged over the
sample. For this averaging procedure one needs the knowledge the field distribution.

As a result of the diffusion vanishing relaxation regime the local inverse
temperatures, $\xi_{mn}(\vec{r},t)$ and $\zeta(\vec{r},t),$ become spatially
distributed over the sample with a distribution which is not the equilibrium
one. In this case we have to take into account also the gradient-terms ,
$\nabla\xi_{mn}\left(  \vec{r},t\right)  $ and $\nabla\zeta\left(  \vec
{r},t\right)  $ in Eqs.(20) and (21). In the next section we will consider the
influence of the spin diffusion process.

\bigskip

3. \textbf{Diffusion relaxation regime}

\bigskip

Assuming that at distances larger then the radius of the diffusion barrier the
diffusion coefficient is independent of $\vec{r}$ \cite{khu66}. Multiplying
Eq. (20) by $\gamma\omega_{mn}^{0}\left(  \vec{r}\right)  \left\langle
e_{mm}\left(  \vec{r}\right)  \right\rangle $ and Eq.(21) by $\sum
_{m}M_{mm\text{ }}$ and by integrating Eqs. (29) and (30) over space variable
$\vec{r}$ \ , we obtain the equations describe the evolution of the
experimentally observed values, the quantaty that conects with the $Z$
component of the total nuclear magnetization ($M_{z}\left(  t\right)
-M_{z}\left(  0\right)  =\sum_{mn}E_{mn}\left(  t\right)  )$, $E_{mn}\left(
t\right)  $%

\begin{align}
\frac{\partial E_{mn}\left(  t\right)  }{\partial t}  &  =-\frac{\gamma}%
{\sum_{m}M_{mm\text{ }}}D_{mn}\int d\vec{r}E_{d}\left(  \vec{r},t\right)
\nabla\omega_{mn}^{0}\left(  \vec{r}\right)  ]\nonumber\\
&  -\int d\vec{r}W_{mn}\left(  \vec{r}\right)  E_{mn}\left(  \vec{r},t\right)
, \tag{29}%
\end{align}
and the total dipolar energy, $E_{d}\left(  t\right)  $%
\begin{equation}
\frac{\partial E_{d}\left(  \vec{r},t\right)  }{\partial t}=-\sum_{mn}%
D_{mn}\int d\vec{r}E_{mn}\left(  \vec{r},t\right)  \Delta\omega_{mn}%
^{0}\left(  \vec{r}\right)  -\int d\vec{r}W_{d}\left(  \vec{r}\right)
E_{d}\left(  \vec{r},t\right)  , \tag{30}%
\end{equation}
where $E_{mn}\left(  t\right)  =\int d\vec{r}E_{mn}\left(  \vec{r},t\right)
$, $E_{mn}\left(  \vec{r},t\right)  =\gamma\omega_{mn}^{0}\left(  \vec
{r}\right)  \xi_{mn}\left\langle e_{mm}\left(  \vec{r},t\right)  \right\rangle
$, $E_{d}\left(  t\right)  =\int d\vec{r}E_{d}\left(  \vec{r},t\right)  $, and
$E_{d}\left(  \vec{r},t\right)  =\sum_{m}M_{mm\text{ }}\zeta\left(  \vec
{r},t\right)  $.

To obtain the solution of the Eqs. (29) and (30) and to calculate the relation
times, both for the nuclear magnetization and for the dipolar energy, we need
to know the internal distribution function of the field, $\omega_{mn}%
^{0}\left(  \vec{r}\right)  $. As it follows from Eqs. (29) and (30) that for
a special distribution of the internal field, $\Delta\omega_{mn}^{0}\left(
\vec{r}\right)  =0$, the diffusion equations gives two uncoupled equations:%

\begin{equation}
\frac{\partial E_{mn}\left(  t\right)  }{\partial t}=-\int d\vec{r}%
W_{mn}\left(  \vec{r}\right)  E_{mn}\left(  \vec{r},t\right)  , \tag{31}%
\end{equation}
and
\begin{equation}
\frac{\partial E_{d}\left(  t\right)  }{\partial t}=-\int d\vec{r}W_{d}\left(
\vec{r}\right)  E_{d}\left(  \vec{r},t\right)  . \tag{32}%
\end{equation}

Solving Eqs. (31) and (32), we obtain the normalized relaxation functions%
\begin{equation}
R_{mn}\left(  t\right)  =\frac{E_{mn}\left(  t\right)  -E_{mn}\left(
\infty\right)  }{E_{mn}\left(  0\right)  -E_{mn}\left(  \infty\right)
}=e^{-t/T_{1mn}} \tag{33}%
\end{equation}
and
\begin{equation}
R_{d}\left(  t\right)  =\frac{E_{d}\left(  t\right)  -E_{d}\left(
\infty\right)  }{E_{d}\left(  0\right)  -E_{d}\left(  \infty\right)
}=e^{-t/T_{1d}}, \tag{34}%
\end{equation}
where $T_{1mn}=\frac{0.12}{C_{p}D_{mn}^{3/4}F}$ and $T_{1d}=\frac{0.12}%
{C_{p}D_{d}^{3/4}F}$, $C_{p}$ is the concentration of the PI and $F$ is the
angular average of the coupling dipolar constant of the DDI between the
nuclear and PI. It follows from the solutions (33) and (34) that at a long
time after the excitation of the spin system, the nuclear magnetization
describes by the sum of exponents $M_{z}\simeq\sum_{mn}a_{mn}e^{-t/T_{1mn}}$,
while the dipolar energy decreases to equilibrium exponentially.

\bigskip

\textbf{Results and discussion}

\bigskip

We will compare the results obtained here with the relaxation processes of the
nuclear magnetization \cite{yu} and the dipolar energy
\cite{GenackRedfield1973} in the mixed state conventional superconducting
vanadium ($I=7/2)$. In the type II superconductors, an applied magnetic field
$\vec{H}_{0},$ in the range between the lower and upper critical field ,
$H_{c1}<H_{0}<H_{c2}$, penetrates into the bulk sample in the form of
vortices, each with a quantum flux of $\Phi_{0}=\frac{c\hbar}{2e}$, which form
a two-dimensional structure in the plane perpendicular to $\vec{H}_{0}$
\cite{A.A.Abrikosov}. The distribution of the internal vortex field can be
obtained by solving the Landau-Ginzburg equation, which gives
\begin{equation}
H\left(  \rho\right)  =\frac{\phi_{0}}{2\pi\lambda^{2}}\ln\frac{\rho}{\lambda
}\text{ for }r<\lambda\tag{35}%
\end{equation}
where $\lambda$ is the London penetration length and $\rho$ is the distance
from the core of vortex in the cylindrical coordinate , $r^{2}=\rho^{2}%
+z^{2}.$ Using experimental data \cite{yu,GenackRedfield1973} we obtain the
spin relaxation time for the nuclear magnetization, $T_{1}=43$ $sec$ and the
spin relaxation time for the dipolar energy, $T_{1d}=93$ $msec$ . As a
consequence, the dipolar energy decreases to the equilibrium state with an
anomalously short time as compare to the relaxation time of the nuclear
magnetization, $\frac{T_{1}}{T_{1d}}=442$. Theoretical estimation of the ratio
of the relaxation time, using Eqs. (33) and (34), results $\frac{T_{1}}%
{T_{1d}}=\left(  \frac{l}{r_{0}}\right)  ^{3}.$Taking into account that the
distance between neihboring vanadium nuclei $r_{0}=2.63\times10^{-8}cm$ and
the radius $l=1.96\times10^{-7}$ $cm$ , for the ratio $\frac{T_{1}}{T_{1d}%
}=413$ , which is in a good agreement with the result obtained from
experimental data.

\bigskip

\textbf{Conclusions}

\bigskip

In conclusion, we obtained coupled equations describing mutual relaxation and
spin diffusion of the quadrupole energy and dipolar energy by using the method
of nonequilibrium state operator \cite{zub}. The equations were solved at a
short and long times approximations corresponding to the direct and diffusion
relaxation regimes. We showed that at the beginning of relaxation process, the
direct relaxation regime is preferred. The relaxation regime changes both for
the nuclear quadrupolar and the dipolar energies, to the diffusion one$.$


\begin{thebibliography}{99}                                                                                               %


\bibitem {bloum}N. Bloembergen, Physica \textbf{15}, 386 (1949).

\bibitem {khu57}G. R. Khutsishvili, Sov. Phys.\ JETP \textbf{4}, 382 (1957).

\bibitem {pdg58}P. G. de Gennes, J.\ Phys.\ Chem.\ Solids \textbf{3}, 345 (1958).

\bibitem {khu66}G. R. Khutsishvili, Sov.\ Phys.\ Uspekhi \textbf{8}, 743 (1966).

\bibitem {khu69}G. R. Khutsishvili, Sov.\ Phys.\ Uspekhi \textbf{11}, 802 (1969).

\bibitem {GenackRedfield1973}A. Z. Genack and A. G. Redfield,
Phys.\ Rev.\ Lett. \textbf{31}, 1204 (1973).

\bibitem {yu}A. G. Redfield and W .N. Yu, Phys.\ Rev. \textbf{169}, 443
(1968), Phys.\ Rev. \textbf{177}, 1018 (1968).

\bibitem {fur1999}G. B. Furman and S. D. Goren, J.Phys.: Condens. Matter,
\textbf{11}, 4045 (1999).

\bibitem {Greenbaum}G.S.Boutis, D. Greenbaum, H. Cho, D.G. Cory, and C.
Ramanathan, Phys. Rev. Lett, \textbf{92}, 137201-1 (2004).

\bibitem {Furman2000}G.B.Furman and S.D.Goren, Sol.St.Nucl.Magn.Res.,
\textbf{16}, 199 (2000).

\bibitem {Furman2004}G.B.Furman and S.D.Goren, Phys. Rev. B 68, 064402 (2004).

\bibitem {jeneer}J. Jeneer and P. Broekaert, Phys.\ Rev. \textbf{157}, 232 (1967).

\bibitem {gold}M. Goldman, \textit{Spin Temperature and Nuclear Resonance in
Solids}(Oxford at the Claredon Press, 1970).

\bibitem {abrg}A. Abragam and M. Goldman, \textit{Nuclear Magnetism: Order and
Disorder}(Clarendon, Oxford,1982) .

\bibitem {slic}C. P. Slichter, \textit{Principles of Magnetic Resonance}%
(Springer-Verlag , Berlin Heidelberg New York, 1980).

\bibitem {bayer6}H.Bayer, Z.Phys., \textbf{130}, 227 (1951).

\bibitem {Furman1983}N.E.Ainbinder and G.B.Furman, Zh. Eksp. Teor. Fiz., 85,
998 (1983) [Sov. Phys. JETP, \textbf{58}, 575 (1983)].

\bibitem {zub}D. N. Zubarev, \textit{Nonequilibrium Statistical
Thermodynamics} (Imprint Consultannts Bureau, New York, 1974).

\bibitem {buizub}L. L. Buishvili and D. N. Zubarev, Sov.\ Phys. \ Solid State
\textbf{3}, 580 (1965).

\bibitem {Furman2000a}G.B.Furman, S.D.Goren, A.M.Panich, and A.I.Shames, Z.
Naturforsch. 55a, pp. 54-60 (2000).

\bibitem {Kubo}K. Kubo and K. Tomita, J. Phys.Soc. Japan, \textbf{9}, 888 (1954).

\bibitem {blum}W. E. Blumberg, Phys.\ Rev., \textbf{119}, 79 (1960).

\bibitem {Ror}H. E. Rorschach Jr., Physica, \textbf{30}, 38 (1964).

\bibitem {das}T.P. Das and E.L. Hahn, Nuclear Quadrupole Resonance
Spectroscopy, Suppl. 1 to Solid State Phys., Academic Press, New York, 1958.

\bibitem {kes}I.L. Bukhbinder and A.R. Kessel, Zh. Eksp. Teor. Fiz., 65, 1498
(1973) [Sov. Phys. JETP, 38, 745, (1974)].

\bibitem {L.D.Landau}L. D. Landau and E. M. Lifshitz, \textit{Quantum
Mechanics -Non Relativistic Theory}, Oxford: Pergamon, 1989.
\end{thebibliography}
\end{document}